\documentclass[12pt]{article}

\usepackage{epsf}
\usepackage{amsmath}
\usepackage{graphics}
\usepackage{amsfonts}
\usepackage{amssymb}
\usepackage{latexsym}
\usepackage{color}
\usepackage[dvips]{graphicx}
\input{colordvi.tex}

\renewcommand{\thefootnote}{\#\arabic{footnote}}
\begin{document}
\setcounter{footnote}{0}

\begin{titlepage}
\begin{flushright}
LMU-ASC 03/12,~RESCEU-2/12
\end{flushright}
\begin{center}

\vskip .5in

{\Large \bf
Fully nonlinear equivalence of $\delta$N and covariant formalisms
}

\vskip .45in

{\large
Teruaki Suyama$^1$,
Yuki Watanabe$^2$
and 
Masahide Yamaguchi$^3$
}

\vskip .45in

{\em
$^1$
  Research Center for the Early Universe (RESCEU), Graduate School
  of Science,\\ The University of Tokyo, Tokyo 113-0033, Japan
  }\\
{\em
$^2$
  Arnold Sommerfeld Center for Theoretical Physics,\\
  Ludwig Maximilian University of Munich, Theresienstrasse 37, 80333 Munich, Germany
  }\\
{\em
$^3$
  Department of Physics, Tokyo Institute of Technology, Tokyo 152-8551, Japan
  }

\end{center}

\vskip .4in

\begin{abstract}
We explicitly show the fully nonlinear equivalence of the $\delta$N and
the covariant formalisms for the superhorizon curvature perturbations,
which enables us to safely evaluate the non-Gaussian quantities of the
curvature perturbation in either formalism. We also discuss
isocurvature perturbations in the covariant formalism and clarify the
relation between the fully nonlinear evolution of the curvature
covector and that of the curvature perturbation for multiple interacting fluids.
\end{abstract}
\end{titlepage}

\renewcommand{\thepage}{\arabic{page}}
\setcounter{page}{1}
\renewcommand{\thefootnote}{\#\arabic{footnote}}

\section{Introduction}

Inflation is now widely accepted as the mechanism of generating primordial
density fluctuations, which are almost scale invariant, adiabatic, and
Gaussian. Such fluctuations are generated inside the horizon and expanded to
superhorizon scales due to the inflationary expansion of the
Universe. These features are confirmed by the recent observations of
cosmic microwave background anisotropies like WMAP experiments
\cite{Komatsu:2010fb}. In particular, the large-scale anticorrelation
of the temperature and the E-mode polarization strongly supports the
presence of the superhorizon epochs of primordial curvature
perturbations \cite{Peiris:2003ff}. Thus, the analysis of the
superhorizon evolution of the curvature perturbations is extremely
important.

Conventionally, the standard perturbative approach has been adopted to
investigate the curvature perturbations \cite{Bardeen:1980kt}. Linear
analysis gives simple equations and solutions, which are used to
evaluate the power spectrum of the curvature perturbations. However,
recent observations are precise enough to probe the nonlinear effects
such as the bispectrum and the trispectrum of the curvature perturbations, which
are crucially important to identify the origin of the primordial density
fluctuations. Though all of the single-field inflation models with the
canonical kinetic term predict negligible nonlinear effects on the
curvature perturbations \cite{Maldacena:2002vr} (see also
\cite{Germani:2011} for a single-field high friction model), inflation
models with noncanonical kinetic terms \cite{Seery:2005wm} and light
field models \cite{Suyama:2010uj} such as curvaton
\cite{Linde:1996gt,Moroi:2002rd} and modulated reheating
\cite{Dvali:2003em,Zaldarriaga:2003my} can predict large non-Gaussianity
of the curvature perturbations. Therefore, the standard perturbative
approach must be extended to deal with the nonlinear effects.  But the
equations and their solutions often become too complicated to be
understood intuitively.  Then, instead of the conventional perturbative
approach, two different formalisms are proposed to deal with the fully
nonlinear curvature perturbations. One is the $\delta$N formalism
\cite{Starobinsky:1986fxa} and the other is the covariant formalism
\cite{Langlois:2005ii,Langlois:2006vv}\footnote{The space gradient of
the fully nonlinear curvature perturbations was first introduced in
\cite{Rigopoulos:2005us} which is based on the leading order of the
coordinate-based gradient expansion. It was later extended to the
curvature covector in a covariant manner by \cite{Langlois:2005ii,
Langlois:2006vv}.}.

The $\delta$N formalism provides a powerful and simple method to
investigate the superhorizon evolution of the fully nonlinear curvature
perturbations. According to the $\delta$N formalism, the superhorizon
curvature perturbation on a uniform energy density hypersurface at a
late time is equal to the perturbation in the time integral of the local
expansion from an initial flat hypersurface to a final uniform energy
density hypersurface. At leading order in gradient expansion, which
corresponds to the separate Universe approach \cite{Wands:2000dp}, each
superhorizon sized region of the Universe evolves in time independently
from other regions. Thus, we have only to evaluate the expansion of the
unperturbed Friedmann Universe. In fact, it is quite easy, by the use
of the $\delta$N formalism, to show that the fully nonlinear curvature
perturbation on a uniform energy density hypersurface is conserved on
superhorizon scales if the pressure is only a function of the energy
density \cite{Lyth:2004gb}.

On the other hand, the covariant formalism defines covariant quantities
corresponding to physical quantities and derives their evolution
equations. Since these quantities are tensor, they are independent of a
particular selection of a coordinate system and easy to understand from
a geometrical perspective. In particular, the fully nonlinear evolution
equations of the curvature covectors corresponding to the curvature
perturbations are easily derived. Notice that, different from the
$\delta$N formalism, they are exact and valid at all scales. Its
covector is shown to be conserved for adiabatic perturbations
\cite{Langlois:2005ii}.

Then, one may wonder what is the relation between the above two
formalisms. In particular, what is the relation between the curvature
perturbation and the curvature covector?  At the linear, second and
third orders \cite{Langlois:2005ii,Enqvist:2006fs,
Lehners:2009ja},\footnote{In Ref. \cite{Enqvist:2006fs}, they
define a scalar quantity similar to the curvature perturbation $\zeta$
and discuss its evolution at all orders.} the curvature covector is
shown to be related to the gradient of the curvature perturbations of
the standard perturbation theory \cite{Malik:2004}. 
For the case of a barotropic fluid, the relation at the fully
nonlinear order has been discussed in \cite{Langlois:2008vk}.
One of the present author (Y.W.)  has evaluated the bispectrum of the curvature
perturbations in two-field inflation models by two different formalisms
and explicitly shown that both of them coincide \cite{Watanabe:2011sm}.
Similar analyses in ekpyrotic models for the bispectrum
\cite{Lehners:2008} and the trispectrum \cite{Lehners:2009ja} have been
done, but they have less accurate quantitative agreement than
\cite{Watanabe:2011sm}.  Nevertheless, to some extent, the $\delta$N
formalism and the covariant formalism is shown to be equivalent for
superhorizon curvature perturbations up to third orders. In this paper,
we explicitly show the fully nonlinear equivalence of the $\delta$N and
the covariant formalisms for the superhorizon curvature perturbations.

The organization of this paper is as follows. In the next two sections,
we briefly review the $\delta$N and the covariant formalisms, and derive
the evolution equations for the curvature perturbation and the curvature
covector from the continuity equation. In Sec.~\ref{sec:equivalence}, we show the fully
nonlinear equivalence of the two formalisms for the superhorizon
curvature perturbations. In Sec.~\ref{sec:isocurvature}, isocurvature perturbations in the
covariant formalism are discussed. We clarify the relation between the
fully nonlinear evolution of the curvature covector and that of the
curvature perturbation for multiple interacting fluids. The final section is devoted to
the summary.

\section{$\delta N$ formalism}\label{sec:deltaN}

In this section, we briefly review the
fully nonlinear version of the $\delta$N formalism, picking up the relevant points derived in \cite{Lyth:2004gb}.

Let us start from the Arnowitt-Deser-Misner (ADM) decomposition of the metric,
\begin{equation}
ds^2=-{\cal N}^2dt^2+\gamma_{ij} (dx^i+\beta^i dt)(dx^j+\beta^j dt). \label{adm-metric}
\end{equation}
We further decompose $\gamma_{ij}$ as 
\begin{equation}
\gamma_{ij}=a^2(t) e^{2 \psi} (e^{h})_{ij}, 
\end{equation}
where $a(t)$ is the fiducial scale factor,
$h_{ij}$ is a traceless tensor, and $\psi$ is the perturbation of a unit spatial volume.

The gradient expansion assumes that all the quantities of interest are
smooth over very large scales \cite{Shibata:1999zs}. Practically, this
expansion is achieved by a prescription that a spatial derivative acting on
any variable is accompanied by a small quantity
$\epsilon$ and we expand all the equations in terms of $\epsilon$.
It leads to the so-called $\delta N$ formalism by keeping only terms at zeroth and first orders in $\epsilon$ and dropping
higher order terms.  As in the
literature, we assume that \footnote{Mathematically speaking, the second
assumption is not needed to derive the evolution equation of $\psi$ in
Eq. (\ref{del-evo-zeta}) but needed to interpret $\psi$ as the spatial curvature.}
\begin{equation}
\beta^i={\cal O}(\epsilon),~~~h'_{ij}={\cal O}(\epsilon^2), \label{impose}
\end{equation}
where the prime ($'$) denotes differentiation with respect to the cosmic time $t$.

Assuming that the anisotropic stress is ${\cal O}(\epsilon^2)$, which is
valid in many cases, we can write the large-scale energy-momentum tensor
of matter filling up the Universe as
\begin{equation}
T_{\mu \nu}=(\rho+P) u_\mu u_\nu+P g_{\mu \nu} + {\cal O}(\epsilon^2).
\end{equation}
The four-vector is normalized to be $u_\mu u^\mu =-1$ and the integration curve along 
$u^\mu$ gives the fluid worldline.
Expansion of $u^\mu$ along the worldline is given by
\begin{equation}
\Theta=\nabla_\mu u^\mu=\frac{3}{\cal N} \left( \frac{a'}{a}+\psi' \right)+{\cal O}(\epsilon^2).
\end{equation}
Correspondingly, the local expansion rate ${\tilde H}$ measured by the local observer 
using his proper time is
\begin{equation}
{\tilde H}=\frac{1}{3} \Theta=\frac{1}{\cal N} \left( \frac{a'}{a}+\psi' \right)+{\cal O}(\epsilon^2).
\end{equation}
The e-folding number of the expansion rate along the worldline is defined as
\begin{equation}
N(t_2,t_1;x^i)=\frac{1}{3} \int_{t_1}^{t_2} dt~{\cal N} \Theta=\log \frac{a(t_2)}{a(t_1)}+\psi (t_2,x^i)-\psi (t_1,x^i)+{\cal O}(\epsilon^2). \label{delta-N}
\end{equation}
Therefore, if the $t={\rm const.}$ hypersurface at initial time $t_1$ is taken to be
a flat one, i.~e.~$\psi(t_1,x^i)=0$, this formula states that perturbation of the 
e-folding number from $t_1$ to $t_2$ coincides with the curvature perturbation 
at $t=t_2$ (up to first order in $\epsilon$). 
Equation (\ref{delta-N}) is the essence of the $\delta N$ formalism.

Time evolution of $\psi$ can be derived from the continuity equation 
$u^\mu \nabla_\nu T^\nu_{~\mu}=0$;
\begin{equation}
\frac{a'(t)}{a(t)}+\psi'(t,x^i)=-\frac{\rho'(t,x^i)}{3\left[\rho(t,x^i)+P(t,x^i)\right]}+{\cal O}(\epsilon^2). \label{del-evo-zeta}
\end{equation}
For a fluid having a barotropic equation of state $P=P(\rho)$, 
we can integrate this equation to obtain a conserved quantity $\zeta$ given by
\begin{equation}\label{eq9}
\zeta (x^i)=\psi(t,x^i)+\frac{1}{3} \int_{\bar \rho(t)}^{\rho(t,x^i)} \frac{d\rho}{\rho+P(\rho)}.
\end{equation}
In particular, if we take a uniform energy density hypersurface
$\Sigma_\rho$, the second term in the right hand side vanishes. Therefore, the
curvature perturbation on that hypersurface remains constant in time.

\section{Covariant formalism}\label{sec:covariant}

In this section, we briefly review the covariant formalism developed in
Refs. \cite{Langlois:2005ii,Langlois:2006vv}.

While the $\delta N$ formalism takes the coordinate-based approach from
the outset, the covariant formalism first defines fully nonlinear and
covariant quantities and then derives the basic equations.  After this,
we can choose a coordinate system and expand the quantities and
equations up to the desired order.

In the covariant formalism, the curvature perturbation is represented by 
a covector $\zeta_a$ defined by
\begin{equation}
\zeta_a = \partial_a N -\frac{\dot N}{\dot \rho} \partial_a \rho=\partial_a N-\frac{\Theta}{3{\dot \rho}} \partial_a \rho.
\end{equation}
We use $a,~b,\cdots$ to stress that we are working in the covariant approach.
Here $N$ is the local e-folding number along the fluid worldline introduced by Eq.~(\ref{delta-N}) 
up to an integration constant that vanishes on some hypersurface.
In the covariant form, it is given by
\begin{equation}
N=\frac{1}{3}\int d\tau~\Theta,
\end{equation}
where $\tau$ is a proper time of the fluid.
The continuity equation $u^a \nabla_b T^b_{~a}=0$ becomes
\begin{equation}\label{eq:continuity}
{\dot \rho}+\Theta (\rho+P)=0,
\end{equation}
where we have assumed that the Universe is filled up with a perfect fluid.
Using this equation, we can derive the evolution equation for $\zeta_a$ along the fluid world-line;
\begin{equation}
{\dot \zeta_a}=-\frac{\Theta}{3(\rho+P)} \left( \partial_a P-\frac{\dot P}{\dot \rho} \partial_a \rho \right). \label{cov-evo-zeta}
\end{equation}
Here the dot (${\dot {}}$) denotes the Lie derivative along the fluid worldline.
For example,
\begin{equation}
{\dot P}=u^a \nabla_a P,~~~~~{\dot \zeta_a}=u^b \partial_b \zeta_a+\zeta_b \partial_a u^b.
\end{equation}
It is worth stressing that the evolution equation (\ref{cov-evo-zeta})
is exact and valid on all length scales.  For fluids having a barotropic
equation of state $P=P(\rho)$, the right-hand side of
Eq.~(\ref{cov-evo-zeta}) vanishes and $\zeta_a$ is therefore conserved
along the worldline.

\section{Equivalence between the two approaches}\label{sec:equivalence}

Now, let us again choose a coordinate system given by
Eq.~(\ref{adm-metric}) with the condition (\ref{impose}) and expand
$\zeta_a$ on this coordinate.  For the time component, we find
\begin{equation}
\zeta_0=N'+\frac{\rho'(t,x^i)}{3\left[\rho(t,x^i)+P(t,x^i)\right]}={\cal O}(\epsilon^2),
\end{equation}
where we have used the continuity equation~(\ref{eq:continuity}) to show the last equality.
For the spatial component, we have 
\begin{equation}
\zeta_i=\partial_i N+\frac{1}{3\left[\rho(t,x^i)+P(t,x^i)\right]} \partial_i \rho(t,x^j).
\end{equation}
Although the right hand side is not generally a total
derivative\footnote{One exception is for the fluid having a barotropic
equation of state $P=P(\rho)$. In \cite{Langlois:2008vk} it has been shown that 
\begin{equation}
\zeta_i=\partial_i\left[N+\frac{\log{\rho}}{3(1+w)}\right],\nonumber
\end{equation}
where $w=P(\rho)/\rho$ is constant in space. Integrating over space and choosing the integration constant properly, one can immediately get
\begin{equation}
\zeta=\delta N + \frac13\int^{\rho}_{\bar\rho(t)}\frac{d\tilde\rho}{(1+w)\tilde\rho},\quad
\delta N\equiv N-\bar{N}(t),\nonumber
\end{equation}
 which is equivalent to Eq.~(\ref{eq9}). In the following, we do not demand the barotropic condition.}, it becomes a total derivative on a
uniform energy density hypersurface $\Sigma_\rho$.  Therefore, we find
that $\zeta_i$ is the spatial derivative of the perturbation of the
e-folding number evaluated on $\Sigma_\rho$.  Using Eq.~(\ref{delta-N}),
$\zeta_i$ can be written as
\begin{equation}
\zeta_i(t_2,x^i)=\partial_i ( \psi (t_2,x^j)-\psi(t_1,x^j))+{\cal O}(\epsilon^3),
\end{equation}
where $\psi (t_2,x^j)$ must be understood as the curvature perturbation
on $\Sigma_\rho$.  In particular, if we take the flat hypersurface at
$t=t_1$, the above equation reduces to
\begin{equation}
\zeta_i(t_2,x^i)=\partial_i \psi (t_2,x^j)+{\cal O}(\epsilon^3).
\end{equation}
At this stage, it is clear that $\zeta_i$ on very large scales is merely
a spatial derivative of the curvature perturbation on $\Sigma_\rho$.

We can explicitly show that the evolution equation (\ref{cov-evo-zeta})
evaluated on $\Sigma_\rho$ indeed leads to Eq.~(\ref{del-evo-zeta}).
The left hand side of Eq.~(\ref{cov-evo-zeta}) is given by
\begin{equation}
{\dot \zeta_i}=\frac{1}{\cal N} \partial_i \psi'+{\cal O}(\epsilon^3).
\end{equation}
On the other hand, the right hand side is given by
\begin{equation}
-\frac{\Theta}{3(\rho+P)} \left( \partial_i P-\frac{\dot P}{\dot \rho} \partial_i \rho \right)=-\frac{\tilde{H}(t,x^i)}{\rho(t)+P(t,x^i)} \partial_i P(t,x^j).
\end{equation}
Then using a formula for ${\cal N}$ valid on $\Sigma_\rho$,
\begin{equation}
{\cal N}=\frac{\rho(t)+P(t)}{\rho(t)+P(t,x^i)},
\end{equation}
and a fact that $\Sigma_\rho$ and the uniform Hubble hypersurface coincide to first order in $\epsilon$ \cite{Lyth:2004gb}, 
we find that Eq.~(\ref{cov-evo-zeta}) reduces to
\begin{equation}
\partial_i \psi'=\partial_i \left[ {H(t)} \frac{\rho(t)+P(t)}{\rho(t)+P(t,x^j)} \right]+{\cal O}(\epsilon^3),
\end{equation}
whose integration over $x^i$, combined with the continuity equation
$\rho'(t)/{\cal N}=-3H(t)\left[\rho(t)\right.+\left.P(t,x^i)\right]$, coincides with
Eq.~(\ref{del-evo-zeta}) provided the integration constant is chosen
properly, i.e., $a'/a$.  The curvature perturbation associated with its
covector obeys the same evolution equations as for the one given in the
$\delta$N formalism, which therefore explicitly establishes the
equivalence at the fully nonlinear level between the $\delta N$
formalism and the covariant formalism.

\section{Isocurvature perturbation in the covariant formalism}\label{sec:isocurvature}

Finally we will show that the evolution equation (\ref{cov-evo-zeta}),
if the fluid consists of multiple interacting fluids, can be written in such a way
that the right-hand side becomes the sum over all the possible combination of the
isocurvature perturbations between two different components, which
resembles the standard one obtained under the linear approximation.
By setting a coordinate system, we clarify the relation between the isocurvature covector and the isocurvature perturbation at the fully nonlinear level.

To rewrite the right hand side of Eq.~(\ref{cov-evo-zeta}) in the desired way, we first introduce the
curvature covector for the fluid component $A$ by
\begin{equation}
\zeta_a^A\equiv\partial_a N-\frac{\Theta}{3 {\dot \rho^A}} \partial_a \rho^A,
\end{equation}
and define the isocurvature covector $S^{AB}_{~a}$ between the fluids A and
B as \cite{Langlois:2006iq}
\begin{equation}
S^{AB}_{~a} \equiv 3 ( \zeta_a^A-\zeta_a^B)=-\Theta \left( \frac{\partial_a \rho^A}{\dot \rho^A}-\frac{\partial_a \rho^B}{\dot \rho^B} \right).
\end{equation}
We can then rewrite the right-hand side of Eq.~(\ref{cov-evo-zeta}) as
\begin{eqnarray}
-\frac{\Theta}{3(\rho+P)} \left( \partial_a P-\frac{\dot P}{\dot \rho} \partial_a \rho \right)=-\frac{\Theta}{3(\rho+P)} \Gamma_a^{\rm (int)}-\frac{\Theta}{6{\dot \rho}^2} \sum_{AB} \left( \frac{\dot P^A}{\dot \rho^A}-\frac{\dot P^B}{\dot \rho^B} \right) {\dot \rho^A} {\dot \rho^B}S^{AB}_{~a},
\end{eqnarray}
where we have assumed that there is no dissipation in total and 
\begin{equation}
\Gamma_a^{\rm (int)} \equiv \sum_A \left( \partial_a P^A-\frac{\dot P^A}{\dot \rho^A} \partial_a \rho^A \right)
\end{equation}
is the sum of the intrinsic nonadiabatic perturbations for each fluid.
Therefore, Eq.~(\ref{cov-evo-zeta}) becomes
\begin{equation}
{\dot \zeta_a}=-\frac{\Theta}{3(\rho+P)} \Gamma_a^{\rm (int)}-\frac{\Theta}{6{\dot \rho}^2} \sum_{AB} \left( \frac{\dot P^A}{\dot \rho^A}-\frac{\dot P^B}{\dot \rho^B} \right) {\dot \rho^A} {\dot \rho^B}S^{AB}_{~a}. \label{iso-evo}
\end{equation}
This result is a fully nonlinear generalization of \cite{Malik:2002jb} and obtained in Ref.~\cite{Langlois:2006iq}. 
One can clearly see
that the evolution of the curvature covector is generated by the
existence of isocurvature covectors  and the relative difference of the sound speeds of fluids.

It is important to
notice that we can apply this equation even when energy is exchanged
among the fluids.  To see it in a more manifest way, let us introduce
the effect of the energy transfer by
\begin{equation}
{\dot \rho^A}+\Theta (\rho^A+P^A)=Q^A, \label{mod-continuity}
\end{equation}
where $Q^A$ represents a rate of energy density that the fluid A
absorbs.
Since the energy is conserved in total, we have $\sum_A Q^A=0$.  
Using this equation, we find that Eq.~(\ref{iso-evo}) can be written as
\begin{eqnarray}
{\dot \zeta_a}&=&-\frac{\Theta}{6} \sum_{AB} 
\left( \frac{\dot P^A}{\dot \rho^A}-\frac{\dot P^B}{\dot \rho^B} \right)
\frac{\rho^A+P^A}{\rho+P} \frac{\rho^B+P^B}{\rho+P}
\left(\frac{\partial_a \rho^A}{\rho^A+P^A}-\frac{\partial_a
 \rho^B}{\rho^B+P^B}\right) \nonumber \\
&&-\frac{\Theta^2}{3 {\dot \rho}^2} 
\biggl( \dot{\rho}\,\Gamma_a^{\rm (int)} + \langle Q \rangle \partial_a \rho
\biggr), \label{iso-evo2} 
\end{eqnarray}
where $\langle Q \rangle \equiv \sum_A c_A^2 Q^A$.  As is clear,
the last term represents a contribution from the energy transfer
but the energy transfer also affects the other terms implicitly
through the evolution equation (\ref{mod-continuity}).

If each fluid obeys a barotropic equation of state [$\Gamma_a^{\rm (int)}=0$], 
by applying the coordinate system given by
Eq.~(\ref{adm-metric}) to Eq.~(\ref{iso-evo2}), the evolution equation
of the curvature perturbation corresponding to Eq.~(\ref{iso-evo2}) can
be derived. Choosing again the uniform energy density hypersurface, the
last term of Eq.~(\ref{iso-evo2}) vanishes and we end up with the
following equation:
\begin{equation}
\partial_i \psi'=-\frac{{\cal N}\Theta}{6} \sum_{AB} 
(c_A^2-c_B^2)
 \frac{\rho^A+P^A}{\rho+P} \frac{\rho^B+P^B}{\rho+P} \partial_i S^{AB},
\end{equation}
where $S^{AB} \equiv 3 (\zeta^A-\zeta^B)$ is the fully nonlinear
isocurvature perturbation introduced in \cite{Kawasaki:2008sn} and
$\zeta^A$ is defined by
\begin{equation}
\zeta^A(t,x^i)\equiv\psi(t,x^i)+\frac{1}{3} \int_{\bar{\rho}^A(t)}^{\rho^A(t,x^i)} \frac{d\rho^A}{\rho^A+P^A(\rho^A)}.
\end{equation}
Thus, we have shown that the evolution of the curvature covector in the
covariant formalism is equivalent to that of the standard curvature
perturbation for multiple interacting fluids as well. Notice that
$\dot P^A/\dot \rho^A$ coincides with the sound velocity squared $c_A^2$
for a barotropic fluid.

\section{Summary}\label{sec:summary}

In this paper, we have shown the fully nonlinear equivalence of
curvature perturbations on superhorizon scales from the two formalisms:
the $\delta N$ and the covariant formalisms.  In particular, by setting
a coordinate system and integrating over space, we have identified the
evolution equation of the curvature covector with that of the curvature
perturbation in the $\delta N$ formalism.  The key assumption here is
that the matter energy-momentum tensor takes the perfect fluid form on
superhorizon scales, i.e., the continuity equation~(\ref{eq:continuity})
completely describes the nonlinear evolution of the energy density of
the system.

We have also clarified the relation between the isocurvature covector
and the nonlinear isocurvature perturbation for multiple interacting
fluids. This identification enables us to bridge the results of standard
perturbation theory to the covariant formalism.  Our treatment here is
somewhat less general than \cite{Langlois:2006iq} but it is enough to
obtain the important relation to the standard curvature perturbation.
It would be interesting to extend our treatment to systems with multiple
scalar, spinor and vector fields, where the next order in gradient
expansion may become important.

\section*{Acknowledgments}

This work was partially supported by the Grant-in-Aid for JSPS Fellows
No.~1008477 (T.S.), the TRR 33 ``The Dark Universe'' (Y.W.) and the Grant-in-Aid for JSPS Scientific Research No.~21740187 (M.Y.).

\end{document}